# Expansions of the solutions of the general Heun equation in terms of the incomplete Beta functions

T.A. Shahverdyan[1], V.M. Red'kov[2], and A.M. Ishkhanyan[1]

[1]Institute for Physical Research of NAS of Armenia, 0203 Ashtarak, Armenia
[2]Institute of Physics of NAS of Belarus, Nezavisimosti av. 68, Minsk 220072, Belarus

Applying the approach based on the equation for the derivative, we construct several expansions of the solutions of the general Heun equation in terms of the incomplete Beta functions. Several expansions in terms of the Appell generalized hypergeometric functions of two variables of the fist kind are also presented. The constructed expansions are applicable for arbitrary sets of the involved parameters. The coefficients of the expansions obey four-, five- or six-term recurrence relations. However, there exist several sets of the parameters for which the recurrence relations involve fewer terms, not necessarily successive. The conditions for deriving finite-sum solutions via termination of the series are discussed.

## 1. Introduction

The general Heun equation and its four confluent reductions are equations with remarkably wide covering in contemporary mathematics, classical and non-classical physics, engineering, chemistry, etc. (see, e.g., [1-5] and references therein). While they suggest a large variety of possible analytically treatable cases for different physical problems that are reduced to the second-order linear ordinary differential equations, the solutions of these equations are expressed in terms of simpler special functions only under rather restrictive circumstances. Such solutions are usually derived via Heun-to-hypergeometric transformations by means of a rational change of the independent variable [6-9], specific quadratures [10] or by termination of infinite series solutions in terms of special functions, in particular, in terms of the functions of the hypergeometric class [11-24].

In the present paper we discuss the series solutions of the general Heun equation in terms of the incomplete Beta functions and the Appell generalized hypergeometric functions. We show that in many cases the constructed expansions provide closed-form solutions achieved via termination of the series. It should be noted, however, that in general the finite sum solutions in terms of incomplete Beta functions are quasi-polynomials, so that the infinite series suggest solutions of much richer structure.

Expansions of the solutions of the general Heun equation in terms of the incomplete Beta functions have been first discussed in [17,18]. Similar expansions have been further proposed for the confluent Heun equation [20]. However, the developed expansions in these cases apply to a rather restricted set of the involved parameters, namely, to the cases when the derivative of the solution of a general or confluent Heun equation is a function of the same class [18,25-27]. This restriction was removed for the confluent Heun equation in [28] where Beta function expansions were constructed for arbitrary parameters involved in the confluent Heun equation. As regards the general Heun equation, in recent papers we have shown that expansions in terms of the incomplete Beta functions can be proposed also for the cases when a characteristic exponent of a finite or infinite singularity of the Heun equation is equal to zero [29,30]. In the present paper we go beyond and show that incomplete Beta function expansions can also be constructed for arbitrary parameters of the general Heun equation. We present two different types of such expansions.

## 2. The first type of expansions

The general Heun equation is written as [1-5]

$$\frac{d^2 u}{dz^2} + \left(\frac{\gamma}{z} + \frac{\delta}{z-1} + \frac{\varepsilon}{z-a}\right)\frac{du}{dz} + \frac{\alpha\beta z - q}{z(z-1)(z-a)} u = 0, \tag{1}$$

where the parameters satisfy the Fuchsian relation $1 + \alpha + \beta = \gamma + \delta + \varepsilon$. This equation has four regular singular points located at $z = 0, 1, a$ and $\infty$. Its solution is denoted as $u = H_G(a, q; \alpha, \beta, \gamma, \delta; z)$ implying that the value of $\varepsilon$ is determined from the Fuchsian relation. Note that below we assume that this notation refers to a solution of Eq. (1) defined up to an arbitrary constant multiplier.

Following the idea of [30], we start with the observation that the function

$$v = z^\gamma (z-1)^\delta \frac{du}{dz} \tag{2}$$

obeys the following second-order differential equation:

$$\frac{d^2 v}{dz^2} + \left(\frac{1-\gamma}{z} + \frac{1-\delta}{z-1} + \frac{1+\varepsilon}{z-a} - \frac{1}{z-z_0}\right)\frac{dv}{dz} + \frac{\Pi(z)}{z(z-1)(z-a)(z-z_0)} v = 0, \tag{3}$$

where $z_0 = q/(\alpha\beta)$ and $\Pi(z)$ denotes the quadratic polynomial

$$\Pi(z) = (\alpha - \varepsilon)(\beta - \varepsilon)z^2 + \left(\gamma\varepsilon - z_0(2\alpha\beta + \varepsilon(1-\alpha-\beta) + \varepsilon^2)\right)z + z_0(z_0\alpha\beta + \varepsilon - \gamma\varepsilon). \tag{4}$$

Equation (3) is a Fuchsian differential equation which in general possesses five regular singular points. Compared with Eq. (1), we have an additional singularity located at the point $z_0 = q/(\alpha\beta)$. Of course, the number of singularities reduces to four if this point coincides with one of the singularities of Eq. (1): $z = 0, 1, a, \infty$. This occurs if $q = 0$, $q = \alpha\beta$, $q = a\alpha\beta$ and $\alpha\beta = 0$, respectively.

The expansions of the solutions of the general Heun equation in terms of the incomplete Beta function and the Appell generalized hypergeometric functions are constructed in the following way. Consider a power-series expansion of a solution of Eq. (3) in the neighborhood of a point $z_1$ of the complex $z$-plane:

$$v = (z - z_1)^\mu \sum_{n=0}^{+\infty} a_n (z - z_1)^n . \qquad (5)$$

The substitution of this series into Eq. (2) and subsequent integration results in the expansion

$$u = C_0 + \sum_{n=0}^{+\infty} a_n \left( \int z^{-\gamma} (z-1)^{-\delta} (z-z_1)^{\mu+n} dz \right). \qquad (6)$$

In general the integrals involved in this expansion are expressed in terms of the Appell generalized hypergeometric functions of two variables of the first kind [4]. In several cases, however, the expansions are written in terms of simpler mathematical functions, in particular, in terms of the incomplete Beta functions. Note that the constant $C_0$ involved in this expansion is a significant component of the expansion; it is not an arbitrary constant. The value of this constant should be particularly specified in order to produce a valid expansion (see examples below).

The choice $z_1 = 0$ produces an expansion in terms of the incomplete Beta functions:

$$u = C_0 + \sum_{n=0}^{+\infty} a_n^{(0)} B(1 + n - \gamma + \mu, 1 - \delta; z), \quad \mu = 0, \gamma, \quad (|z| \leq 1). \qquad (7)$$

Substituting this expansion into Eq. (1) and taking the limit $z \to 0$ we readily find that here $C_0 = a\mu/q$ if $\operatorname{Re}(1 - \gamma + \mu) > 0$.

Similarly, choosing $z_1 = 1$ we get another expansion involving incomplete Beta functions with interchanged parameters as compared with the previous case:

$$u = C_0 + \sum_{n=0}^{+\infty} a_n^{(1)} (-1)^{n+\mu-\delta} B(1 - \gamma, 1 + n - \delta + \mu; z), \quad \mu = 0, \delta, \qquad (8)$$

where the constant $C_0$ should be calculated separately.

Above two expansions extend to the case of arbitrary nonzero $\varepsilon$ two of the expansions of [30] written in terms of exactly the same incomplete Beta functions as those involved in Eqs. (7) and (8). The coefficients of the derived expansions obey four-term recurrence relations. For instance, for the coefficients $a_n^{(0)}$ involved in Eq. (7) we have:

$$S_n a_n^{(0)} + R_{n-1} a_{n-1}^{(0)} + Q_{n-2} a_{n-2}^{(0)} + P_{n-3} a_{n-3}^{(0)} = 0, \qquad (9)$$

where

$$S_n = -a z_0 (n + \mu)(n - \gamma + \mu), \qquad (10)$$

$$R_n = r_0 + r_1 (n + \mu) + r_2 (n + \mu)^2, \quad Q_n = q_0 + q_1 (n + \mu) + q_2 (n + \mu)^2, \qquad (11)$$

$$P_n = (n - \alpha + \varepsilon + \mu)(n - \beta + \varepsilon + \mu), \qquad (12)$$

where the parameters $r_{0,1,2}$, $q_{0,1,2}$ do not depend on $n$. It is seen that the recurrence relation becomes three-term if $z_0 = 0$, that is if $q = 0$. If $q \neq 0$, the series is left-hand side terminated at $n = 0$ if $S_0 = 0$, i.e., if $\mu = 0$ or $\mu = \gamma$. It will terminate from the right-hand side if three successive coefficients vanish for some $N = 1, 2, \ldots$: $a_N \neq 0$, $a_{N+1} = a_{N+2} = a_{N+3} = 0$. From the equation $a_{N+3} = 0$ we find that the termination is possible if $P_N = 0$, i.e., if

$$\alpha = N + \varepsilon + \mu \text{ or } \beta = N + \varepsilon + \mu, \quad \mu = 0, \gamma. \qquad (13)$$

The remaining two equations, $a_{N+1} = 0$ and $a_{N+2} = 0$, impose two more restrictions on the parameters of the Heun equation.

As already stated above, in the general case $z_1 \neq 0,1$ Eq. (6) presents an expansion in terms of the Appell generalized hypergeometric functions of the first kind:

$$u = C_0 + \sum_{n=0}^{+\infty} a_n^{(z_1)} \frac{(-1)^{-\delta}}{(-z_1)^{-\mu-n}} \frac{z^{1-\gamma}}{1-\gamma} F_1\left(1 - \gamma;\ \delta,\ -\mu - n;\ 2 - \gamma;\ z, \frac{z}{z_1}\right). \qquad (14)$$

It follows from Eq. (3) that $\mu = 0$ or $-\varepsilon$ if $z_1 = a$ and $\mu = 0$ or $2$ if $z_1 = z_0 = q/(\alpha\beta)$. In several cases the Appell functions of this expansion are also reduced to incomplete Beta functions. For instance, this occurs if $\delta$ or $\gamma$ is a non-positive integer. For $\delta = 0$ we have

$$u(\delta = 0) = C_0 + \sum_{n=0}^{+\infty} a_n^{(z_1)} (-z_1)^n B\left(1 - \gamma,\ 1 + n + \mu;\ \frac{z}{z_1}\right), \qquad (15)$$

and for $\gamma = 0$ the result reads

$$u(\gamma = 0) = C_0 + \sum_{n=0}^{+\infty} a_n^{(z_1)} (1 - z_1)^n B\left(1 + n + \mu,\ 1 - \delta;\ \frac{z - z_1}{1 - z_1}\right). \qquad (16)$$

The last two expansions have also been presented in [30].

The coefficients $a_n^{(z_1)}$ of expansion (14) also obey a four-term recurrence relation. For instance, for $z_1 = z_0 = q/(\alpha\beta)$ we have

$$S_n a_n^{(z_0)} + R_{n-1} a_{n-1}^{(z_0)} + Q_{n-2} a_{n-2}^{(z_0)} + P_{n-3} a_{n-3}^{(z_0)} = 0, \tag{17}$$

with

$$S_n = z_0 (z_0 - 1)(z_0 - a)(n + \mu)(n - 2 + \mu), \tag{18}$$

$$R_n = r_0 + r_1 (n + \mu) + r_2 (n + \mu)^2, \quad Q_n = q_0 + q_1 (n + \mu) + q_2 (n + \mu)^2, \tag{19}$$

$$P_n = (n - \alpha + \varepsilon + \mu)(n - \beta + \varepsilon + \mu), \tag{20}$$

where again the parameters $r_{0,1,2}$, $q_{0,1,2}$ do not depend on $n$. This recurrence relation involves just three successive terms only if $q = 0$, $q = \alpha\beta$, or $q = a\alpha\beta$, that is if $z_0$ coincides with one of the existing singular points of the general Heun equation. Except for these three specific cases, the series may left-hand side terminate if $\mu = 0$ or $\mu = 2$. Since the exponents differ by an integer, we may expect to have only one consistent power-series expansion corresponding to the greater exponent $\mu = 2$; the second solution may require a logarithmic term. The series with $\mu = 2$ may terminate from the right-hand side at some $N = 1, 2, \ldots$ if $P_N = 0$, or, equivalently, if

$$\alpha = N + \varepsilon + 2 \text{ or } \beta = N + \varepsilon + 2. \tag{21}$$

The termination will occur if additionally $a_{N+1} = 0$ and $a_{N+2} = 0$, which impose two more restrictions on the involved parameters.

Under some restrictions, the recurrence relations (9) and (17) may involve just two terms. For instance, the coefficients $R_n$ and $Q_n$ simultaneously vanish in Eq. (9) for all $n$ if

$$a = (-1)^{\pm 2/3}, \quad q = \pm 2i/\sqrt{3}, \quad \alpha = 1, \quad \beta = -1 - q/2, \quad \gamma = 1 - q/2, \quad \delta = 2, \quad \varepsilon = -2. \tag{22}$$

In all these cases the recurrence equation is explicitly solved and the solution of the Heun equation is written in terms of hypergeometric functions. An example is the case

$$a = 1/2, \quad q = a\alpha\beta, \quad \gamma = \delta, \tag{23}$$

when the coefficients $S_n$ and $Q_n$ in Eq. (17) simultaneously vanish for all $n$, so that the recurrence relation simplifies to $R_{n-1} a_{n-1}^{(0)} + P_{n-3} a_{n-3}^{(0)} = 0$. The solution of the Heun equation in then calculated to be

$$u = C_1 \, {}_2F_1\left(\frac{\gamma - 1 - r}{4}, \frac{\gamma - 1 + r}{4}; \frac{1 + \gamma}{2}; z^2\right) + C_2 z^{1-\gamma} \, {}_2F_1\left(\frac{1 - \gamma - r}{4}, \frac{1 - \gamma + r}{4}; \frac{3 - \gamma}{2}; z^2\right), \tag{24}$$

where $r = \sqrt{(\gamma - 1)^2 - 4\alpha\beta}$.

## 3. The second type of expansions

Other expansions in terms of the incomplete Beta functions or in terms of the Appell generalized hypergeometric functions can be constructed, e.g., if a preliminary change of the dependent variable is applied. For instance, applying the transformation $u = (z-a)^{-\varepsilon/2} w(z)$, Eq. (1) is rewritten as

$$\frac{d^2w}{dz^2} + \left(\frac{\gamma}{z} + \frac{\delta}{z-1}\right)\frac{dw}{dz} + \frac{\Pi(z)}{z(z-1)(z-a)^2} w = 0, \qquad (25)$$

where $\Pi(z)$ denotes the following polynomial:

$$\Pi(z) = \left(\alpha - \frac{\varepsilon}{2}\right)\left(\beta - \frac{\varepsilon}{2}\right)z^2 + \left(\frac{\varepsilon^2}{4} + \frac{\varepsilon}{2}(\gamma - 1 + a\gamma + a\delta) - (q + a\alpha\beta)\right)z + a\left(q - \frac{\gamma\varepsilon}{2}\right). \qquad (26)$$

If $\varepsilon \neq 2\alpha$ or $2\beta$, this is a quadratic polynomial that can be factorized as $p_0(z - z_1)(z - z_2)$, where $p_0 = (\alpha - \varepsilon/2)(\beta - \varepsilon/2)$. It is then readily verified that the function

$$v = z^\gamma (z-1)^\delta \frac{dw}{dz} \qquad (27)$$

obeys the following second-order equation:

$$\frac{d^2v}{dz^2} + \left(\frac{1-\gamma}{z} + \frac{1-\delta}{z-1} + \frac{2}{z-a} - \frac{1}{z-z_1} - \frac{1}{z-z_2}\right)\frac{dv}{dz} + \frac{p_0(z-z_1)(z-z_2)}{z(z-1)(z-a)^2} v = 0. \qquad (28)$$

In general, this equation has six regular singular points; we have two additional singularities located at $z = z_1$ and $z = z_2$. Of course, the overall number of singularities decreases if one or both of these new singularities coincide with already existing singularities $z = 0, 1, a$.

If $\varepsilon = 2\alpha$ or $2\beta$, $\Pi(z)$ becomes a linear function of $z$: $\Pi(z) = p_1(z - z_1)$, and the transformation (27) now produces an equation having five singular points:

$$\frac{d^2v}{dz^2} + \left(\frac{1-\gamma}{z} + \frac{1-\delta}{z-1} + \frac{2}{z-a} - \frac{1}{z-z_1}\right)\frac{dv}{dz} + \frac{p_1(z-z_1)}{z(z-1)(z-a)^2} v = 0. \qquad (29)$$

Now, again considering a power-series expansion of a solution of Eq. (28) or Eq. (29) in the neighborhood of a point $z_i$ of the complex $z$-plane:

$$v = (z - z_i)^\mu \sum_{n=0}^{+\infty} a_n (z - z_i)^n, \qquad (30)$$

we arrive at the expansion (compare with Eq. (6))

$$u = (z-a)^{-\varepsilon/2}\left[C_0 + \sum_{n=0}^{+\infty} a_n \left(\int z^{-\gamma}(z-1)^{-\delta}(z-z_i)^{\mu+n} dz\right)\right]. \qquad (31)$$

The integrals here are of the same form as those of Eq. (6), hence, similar developments as the above expansions (7), (8) and (14) apply.

Thus, the choice $z_i = 0$ produces an expansion in terms of incomplete Beta functions:

$$u = (z-a)^{-\varepsilon/2}\left[C_0 + \sum_{n=0}^{+\infty} a_n^{(0)} B(1+n-\gamma+\mu, 1-\delta; z)\right], \quad \mu = 0, \gamma, \tag{32}$$

where now $C_0 = a\mu/(q - \gamma\varepsilon/2)$ if $\text{Re}(1-\gamma+\mu) > 0$.

Similarly, the choice $z_i = 1$ produces another incomplete Beta function expansion:

$$u = (z-a)^{-\varepsilon/2}\left[C_0 + \sum_{n=0}^{+\infty} a_n^{(1)}(-1)^{n+\mu} B(1-\gamma, 1+n-\delta+\mu; z)\right], \quad \mu = 0, \delta. \tag{33}$$

And finally, in the general case $z_i \neq 0, 1$ we have an expansion in terms of the Appell generalized hypergeometric functions of the first kind:

$$u = (z-a)^{-\varepsilon/2}\left[C_0 + \sum_{n=0}^{+\infty} a_n^{(z_i)} \frac{(-1)^{-\delta}}{(-z_i)^{-\mu-n}} \frac{z^{1-\gamma}}{1-\gamma} F_1\left(1-\gamma; \delta, -\mu-n; 2-\gamma; z, \frac{z}{z_i}\right)\right]. \tag{34}$$

Again, as in the case of expansion (14), in several cases the Appell functions are reduced to incomplete Beta functions, e.g., if $\gamma$ or $\delta$ is zero or a negative integer (see Eqs. (15), (16)).

The first two of these expansions, Eqs. (32) and (33), present different extensions to the case of nonzero $\varepsilon$ of two of the expansions presented in [30], which are written in terms of the same incomplete Beta functions as those involved in Eqs. (32) and (33).

Though the above two types of expansions, Eqs. (7)-(8),(14) and (32)-(34), have much in common, there is a difference between the expansions of the first and the second types that is worth to mentioning. This concerns the recurrence relations between the coefficients of the expansions. For the first type expansions these are four-term relations, while for the second type expansions the relations are in general six-term that are reduced to five- or four-term ones only under rather restrictive conditions.

Indeed, consider the coefficients $a_n^{(0)}$ of expansion (32). In the general case $\varepsilon \neq 2\alpha$, $2\beta$ the equation obeyed by $v(z)$ is Eq. (28). This leads to a six-term recurrence relation:

$$K_n a_n^{(0)} + T_{n-1} a_{n-1}^{(0)} + S_{n-2} a_{n-2}^{(0)} + R_{n-3} a_{n-3}^{(0)} + Q_{n-4} a_{n-4}^{(0)} + + P_{n-5} a_{n-5}^{(0)} = 0. \tag{35}$$

Though the coefficients of this relation are cumbersome, however, they are informative and useful for several general observations. We here write in explicit form just two of them:

$$K_n = -a^2 z_1 z_2 (\mu+n)(\mu+n-\gamma), \tag{36}$$

$$P_n = (\mu+n-\alpha+\varepsilon/2)(\mu+n-\beta+\varepsilon/2), \tag{37}$$

Apart from the trivial case $a = 0$, the recurrence relation (35) may reduce to one involving five successive terms only if $z_{1,2} = 0$. If none of $a, z_1, z_2$ is zero, the series is left-hand side terminated at $n = 0$ if $K_0 = 0$, i.e., if $\mu = 0$ or $\mu = \gamma$. For the limit $\rho = \lim_{n\to\infty}\left(a_n^{(0)}/a_{n-1}^{(0)}\right)$ a quintic polynomial equation holds, the roots of which are 1, $1/a$, $1/a$, $1/z_1$, $1/z_2$. Unless conditioned for augmented convergence, $a_n/a_{n-1}$ tends to the larger root $\rho_{max}$ so that the series (30) is convergent if $|z\rho_{max}| < 1$. Hence, the convergence radius of the expansion (30) is $\min\{1, |a|, |z_1|, |z_2|\}$. From Eq. (37) we see that the series may terminate from the right-hand side at some $N = 1, 2, \ldots$ if

$$\alpha = N + \mu + \varepsilon/2 \text{ or } \beta = N + \mu + \varepsilon/2. \tag{38}$$

The termination will actually occur if additionally $a_{N+1} = a_{N+2} = a_{N+3} = a_{N+4} = 0$, hence, in this case four more restrictions are to be imposed on the parameters of the Heun equation.

If $\varepsilon = 2\alpha$ or $2\beta$, the proper equation for the function $v(z)$ is Eq. (29). This results in a five-term recurrence relation:

$$T_n a_n^{(0)} + S_{n-1} a_{n-1}^{(0)} + R_{n-2} a_{n-2}^{(0)} + Q_{n-3} a_{n-3}^{(0)} + P_{n-4} a_{n-4}^{(0)} = 0, \tag{39}$$

the coefficients of which are explicitly written as

$$T_n = a^2 z_1 (\mu + n)(\mu + n - \gamma), \tag{40}$$

$$S_n = p_1 z_1^2 + a(a(1+\gamma) + 2z_1(\gamma - 1) + az_1(\gamma + \delta - 1))\mu_1 - a(2z_1 + a(1+z_1))\mu_1^2 \tag{41}$$

$$R_n = (a^2 + 2a(z_1+1) + z_1)\mu_1^2 - (2a\gamma + a^2(\gamma + \delta) + 2az_1(\gamma + \delta - 2) + (\gamma - 2)z_1)\mu_1 - 2p_1 z_1, \tag{42}$$

$$Q_n = p_1 + (\gamma - 1 + 2a(\gamma + \delta - 1) + z_1(\gamma + \delta - 3))\mu_1 - (2a + z_1 + 1)\mu_1^2, \tag{43}$$

$$P_n = (\mu + n)(\mu + n + 2 - \gamma - \delta), \tag{44}$$

where $\mu_1 = \mu + n$. Apart from the trivial case $a = 0$, the recurrence relation (39) may reduce to one involving only four successive terms if $z_1 = 0$, i.e., $q = \gamma\varepsilon/2 = \gamma\alpha$. If none of $a, z_1$ is zero, the series is left-hand side terminated at $n = 0$ if $\mu = 0$ or $\mu = \gamma$. For the limit $\rho = \lim_{n\to\infty}\left(a_n^{(0)}/a_{n-1}^{(0)}\right)$ a quartic polynomial equation holds, the roots of which are 1, $1/a$, $1/a$, $1/z_1$, so that the convergence radius of the expansion (27) this time is $\min\{1, |a|, |z_1|\}$. The series may terminate from the right-hand side at some $N = 1, 2, \ldots$ when

$$\gamma + \delta = N + 2 \text{ if } \mu = 0 \tag{45}$$

and

$$\gamma = -N \text{ or } \delta = N + 2 \text{ if } \mu = \gamma. \tag{46}$$

It should additionally be $a_{N+1} = a_{N+2} = a_{N+3} = 0$, hence, for finite-sum solutions in this case three more restrictions are to be imposed on the parameters of the Heun equation.

**4. Discussion**

Thus, using an equation obeyed by a function proportional to the derivative of a solution of the Heun equation, we have constructed several expansions of the solutions of the general Heun equation in terms of the incomplete Beta functions and the Appell generalized hypergeometric functions. The expansions apply to arbitrary sets of the involved parameters. The coefficients of the expansions in general obey four-term recurrence relations, which are reduced to the ones involving fewer terms under some restrictions imposed on the parameters of the Heun equation. We have presented some cases governed by two-term recurrence relations when the corresponding series are explicitly summed and the resultant solution of the Heun equation is written in terms of hypergeometric functions.

We have shown that other expansions in terms of the incomplete Beta functions and the Appell generalized hypergeometric functions can be constructed if a preliminary change of the dependent and/or independent variables is applied. However, in this case the recurrence relations for the coefficients of the expansions may involve more terms. In the particular case presented above these relations are five- or six-term. It would be interesting to explore other possibilities. For instance, to see if a transformation of the form $u(z) = z^{\gamma_0} (z-1)^{\delta_0} (z-a)^{\varepsilon_0} w(z)$, with adjustable parameters $\gamma_0, \delta_0, \varepsilon_0$, is able to produce simpler recurrence relations for some sets of the involved parameters, or if such a transformation, perhaps combined with a rational change of the independent variable, can produce new explicit solutions in terms of simpler mathematical functions.

A final remark is that the above approach based on the equation for the derivative can be used to construct expansions in terms of other higher transcendental functions, e.g., the Goursat generalized hypergeometric functions [26]. Further possibilities for such a kind of expansions are suggested if integral representations, equations and relations for the Heun functions are applied [31-35].

**Acknowledgments**

This research has been conducted within the scope of the International Associated Laboratory (CNRS-France & SCS-Armenia) IRMAS. The work has been supported by the Armenian State Committee of Science (SCS Grants No. 13RB-052 and No. 15T-1C323).